\documentclass[11pt]{iopart}
\usepackage{iopams}
\usepackage{txfonts}
\usepackage{graphicx}
\begin{document}

\article[Modelling Starbursts in HII Galaxies]{Star-forming Dwarf Galaxies: Ariadne's Thread in the
  Cosmic Labyrinth}{Modelling Starbursts in
HII Galaxies: What do we need to fit the observations?}

\author{Mart\' {i}n-Manj\'{o}n, M.L. $^{1}$, Moll\' {a}, M. $^{2}$,
D\' {i}az, A.I. $^{1}$ \& Terlevich, R.$^{3}$}

\address{1. Universidad Autónoma de Madrid, Madrid (Spain),
2. CIEMAT, Madrid (Spain), 
3. INAOE, Puebla (Mexico)}
\ead{mariluz.martin@uam.es}
\begin{abstract}

We have computed a series of realistic and self-consistent models that
have been shown to be able to reproduce the emitted spectra of HII
galaxies in a star bursting scenario. Our models combine different
codes of chemical evolution, evolutionary population synthesis and
photoionization. The emitted spectrum of HII galaxies is reproduced by
means of the photoionization code CLOUDY \cite{fer98}, using as
ionizing spectrum the spectral energy distribution (SED) of the
modelled HII galaxy, calculated using the new and updated stellar
population models PopStar (\cite{mol08}, in prep.).This, in turn, is
calculated according to a star formation history and a metallicity
evolution given by a chemical evolution model. Each model is
characterized by three parameters which are going to determine the
evolution of the modeled galaxy: an initial efficiency of star formation, the
way in which burst take place, and the time of separation between
these bursts. Some model results emerging from the combination of
different values for these three parameters are shown here. Our
technique reproduces observed abundances, diagnostic diagrams and
equivalent width-colour relations for local HII galaxies.

\end{abstract}

\maketitle

\section{Introduction}

 HII galaxies are characterized by strong and narrow emission lines
 and by a low metal content, but this does not necessarily mean that
 these galaxies be young systems. The current burst of star formation
 (SF) dominates the SED even if previous stellar populations are
 present, making difficult to know the star formation history (SFH) of
 the galaxy. We have made a grid of attenuated star-bursting models,
 based on \cite{mmm08a}, using simultaneously the whole information
 available for the galaxy sample: the ionized gas, which defines the
 present time state of the galaxy, and spectrophotometric parameters,
 related to its SFH. The models are computed in a self-consistent way,
 that is using the same assumptions regarding stellar evolution, model
 stellar atmospheres and nucleosynthesis, and a realistic
 age-metallicity relation.

\section{The Star-Bursting Model}

The model consists in a set of successive instantaneous bursts of star
formation in a region with a total mass of gas of 100$\cdot$10$^{6}$
M$_{\odot}$, which take place along the whole evolution of the galaxy
in 13.2 Gyr.  The\textbf{ chemical evolution code} used is based on
\cite{mol05}. We obtain for every 0.7 Myr time step the abundances of
15 elements: H, He, C, O, N, Ne, Na, Al, Mg, Si, S, Ca, Ar, Ni, Fe,
the star formation rate (SFR) and the corresponding age-metallicity
relation, Z(t). With this SFH and Z(t) we assign a SED from the
library \textbf{Popstar} \cite[in prep.]{mol08}, with a Ferrini IMF to
each time step stellar population. When more than one burst takes
place the resulting SED is the sum of the SEDs of every stellar
generation convolved with the SFH. The final result is the total
luminosity at each wavelength of the whole stellar population,
including the ionizing continuum of the last formed stellar
generation.  This resulting SED is used as ionizing source for the
photoionization code \textbf{CLOUDY} \cite{fer98}, which gives the
emission lines produced by the modelled nebula. The gas is ionized by
the massive stars of the current burst of SF, which is characterized
by a radius R, calculated according to the mechanical energy output of
the massive stars winds and SNeI explosions, a gas density, n$_{H}$,
the number of Lyman ionizing photons Q(H), obtained directly from the
SEDs of the ionizing continuum, and the chemical abundances obtained
from the chemical evolution code.


Each model is characterized by three input parameters:

\begin{itemize}
\item \textbf{The initial efficiency (\textit{$\epsilon$})}: It is the
amount of gas consumed to form stars in the first burst of star
formation. We present here the models made with the
percentages of 33$\%$ and 10$\%$ , that is, in these models, the first
burst of star formation involves 33$\cdot$10$^{6}$ M$_{\odot}$
(\textbf{high efficiency model}), and 10$\cdot$10$^{6}$ M$_{\odot}$
(\textbf{low efficiency model}), respectively.

\item \textbf{Attenuation (\textit{k)}}: The initial efficiency of
star formation(SF) is attenuated in two different ways:
\begin{enumerate}
\item By a factor which changes with the number of the burst,
\textbf{\textit{n}}, following the expression:

\[
\Psi_{n}=(\frac{1}{n})\cdot\Psi_{0} 
\]

which corresponds to a \textbf{soft attenuation}.

\item By a constant factor, \textbf{\textit{k$^{(n-1)}$}}, according
to the expression:

\[
\Psi_{n}=\Psi_{0}\cdot k^{(n-1)}
\]

where \textbf{\textit{n}} is the number of the current burst and
\textbf{\textit{k}} the attenuation factor. For this work we have
taken \textbf{k=0.65}, corresponding to a \textbf{strong attenuation}.
\end{enumerate}

\item \textbf{Time between bursts (\textbf{\textit{$\Delta$t}})}:
Every burst takes place instantaneously and it is followed by quiet
periods, whose duration can change. For this work we have taken
$\Delta$t= 1.3 Gyr for the inter-burst time, that is, one burst every
1.3 Gyr. For comparison purposes, we are going to show some results of
models with $\Delta$t=0.1 Gyr and $\Delta$t=0.05 Gyr.

\end{itemize}

\section{Results}

\begin{figure}[!ht]
\includegraphics[clip,width=0.5\textwidth,angle=0]{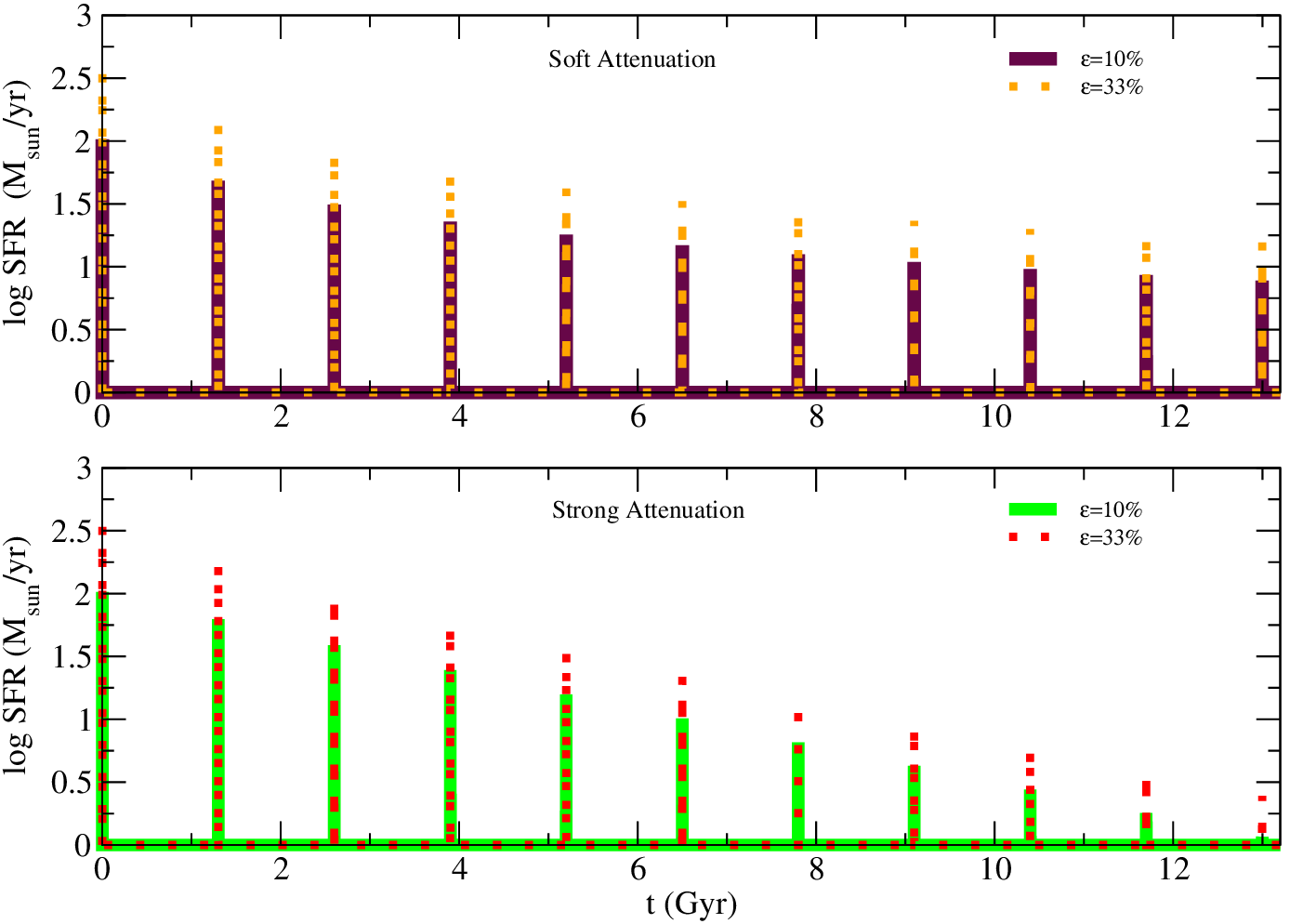}
\includegraphics[clip,width=0.5\textwidth,angle=0]{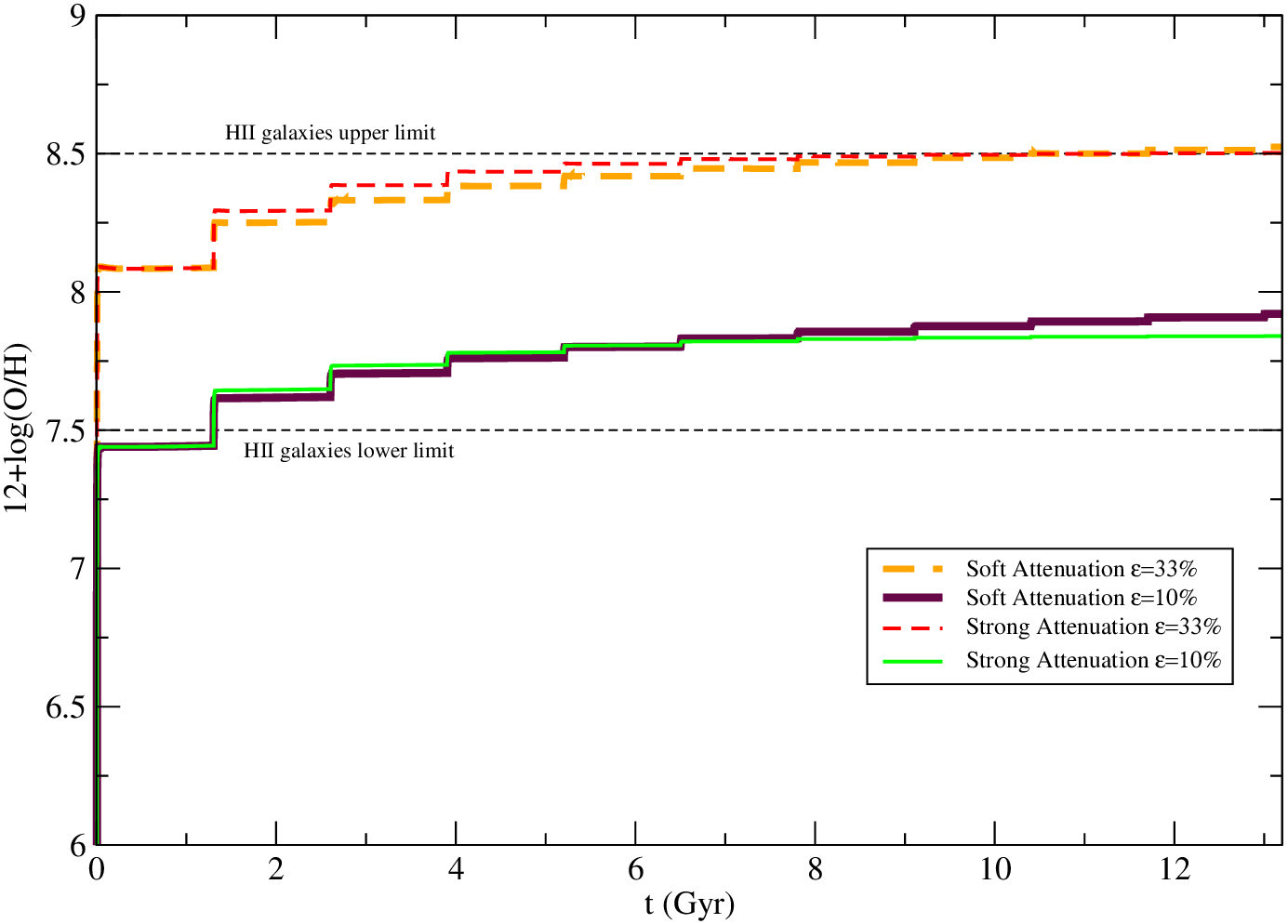}
\caption{Left: SFR of the models with soft (upper
part) and strong attenuation (lower part). Both efficiencies are
showed in each panel. Right: Evolution of oxygen abundance for models
with different initial efficiencies and
attenuation values.}
\label{fig1}
\end{figure}

\textbf{The initial efficiency of the star formation principally leads
the star formation rate and the initial oxygen abundance}. The SFR and
oxygen abundances of our models are between the values observed in HII
galaxies \cite{hoy04, hoy06}. In figure \ref{fig1}, left panel, we can
see that the first burst is strong, while the subsequent ones are less
intense due to the decrease of the available gas and attenuation. The
two efficiencies chosen, 33$\%$ and 10$\%$, give the upper and lower
limits respectively for HII galaxy oxygen abundance range, as can be
seen in figure \ref{fig1}, right panel. Models with the same initial
efficiency, but different attenuation type, are very similar.

\begin{figure}[!ht]
\includegraphics[clip,width=0.5\textwidth,angle=0]{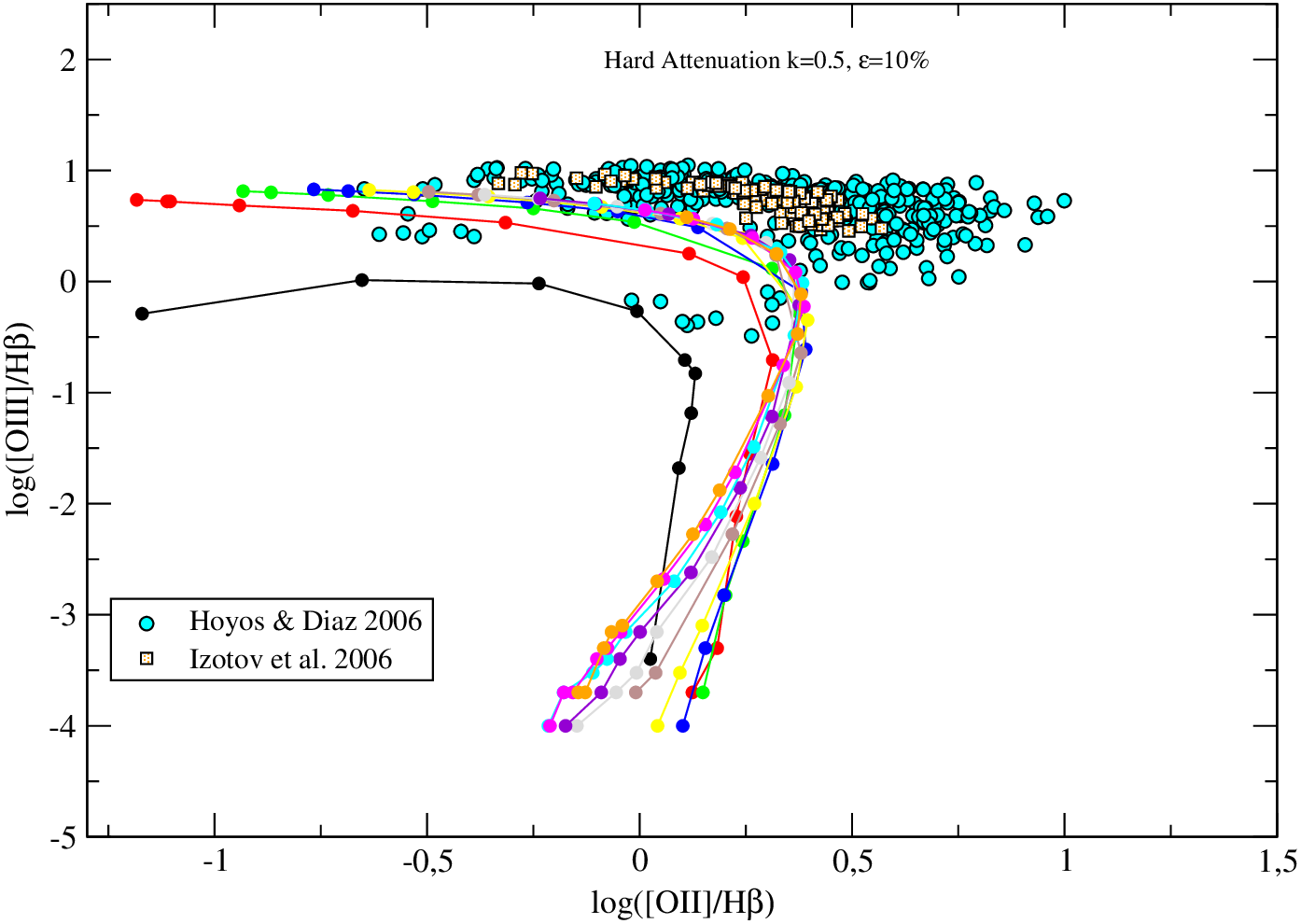}
\includegraphics[clip,width=0.5\textwidth,angle=0]{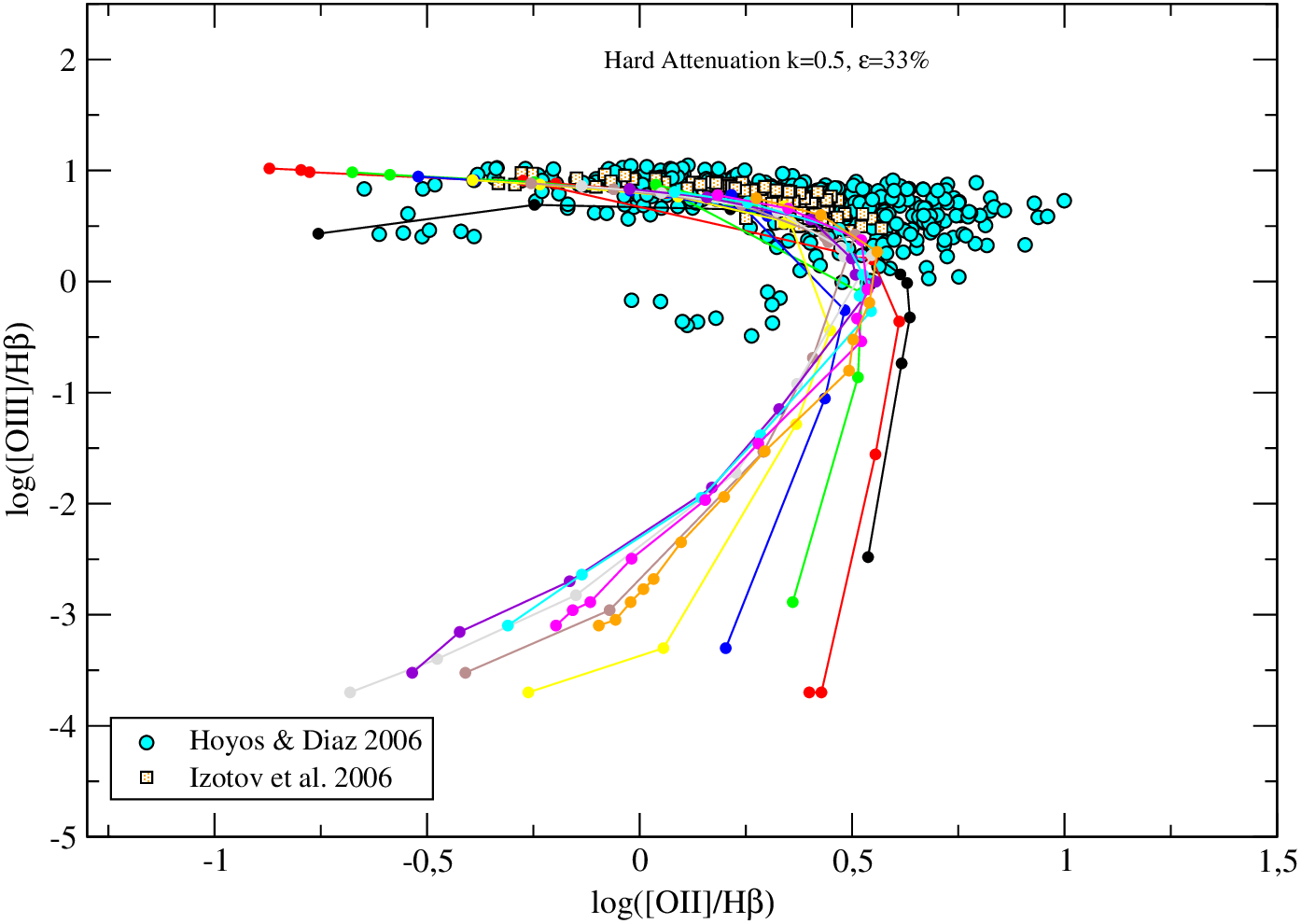}
\caption{Diagnostic diagram for the Strong
Attenuation Model with the low efficiency case at the left and the
high efficiency case at the right.  Observational data from
\cite{hoy06, izo06}. The different coloured lines represent each SF burst, from the first one ocurred at t=0 Gyr (black line) to the last one at t=13 Gyr (violet line).}
\label{fig2}
\end{figure}

The initial efficiency also leads the behaviour of the ionized
gas. The emission lines are produced by the ionizing photons of the
massive stars present in the current burst. In figures \ref{fig2} we
can see the differences between the models with a high initial
efficiency (33$\%$), right panel, which reproduce high excitation and
high abundance galaxies, with high [OIII]$\lambda$5007/H$_{\beta}$,
due to its high efficiency of SF, and those with low efficiency (
10$\%$),left panel, which reproduce less metallic galaxies, with high
[OIII]$\lambda$5007/H$_{\beta}$ and low
[OIII]$\lambda$5007/H$_{\beta}$ ratios. Differences due to attenuation
are not so important since the SFR of both models are very similar.

\begin{figure}
\begin{center}
\includegraphics[clip,width=0.9\textwidth,angle=0]{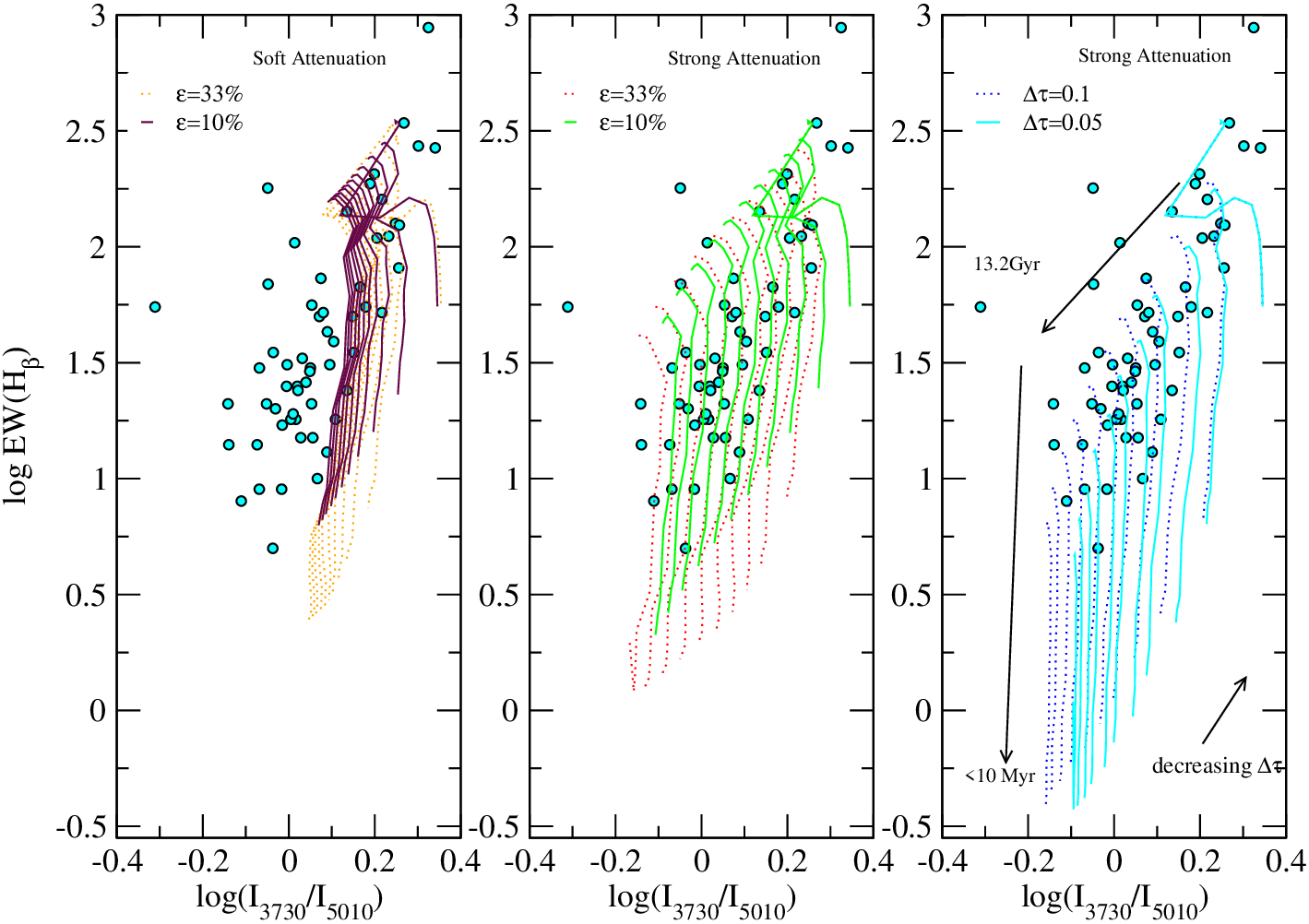}
\caption{ \begin{footnotesize}EW(H$_{\beta}$)
vs. log(I$_{3730}$/I$_{5010}$) for different attenuation 
parameters: a soft attenuation model (left panel), a strong attenuation
model (central panel), and a strong attenuation model with different
inter-burst times (right panel). Both initial efficiencies are
represented in the two first panels and in the last one, only 33\% models for different inter-burst times are represented.\end{footnotesize}}
\label{fig3}
\end{center}
\end{figure}

\textbf{The attenuation of the bursts (k) determines the contribution
of the underlying population:} The SFR for the successive bursts and
the oxygen abundance evolution are set by the attenuation too, keeping
them within the range of the observations. Furthermore, a higher
attenuation implies a larger contribution from the previous bursts to
the total SED. Then, the most important characteristics given by the
adjustment of the attenuation are the colours of the continuum and the
evolution of the equivalent width of H$_{\beta}$. The evolution of
EW(H$_{\beta}$) {\sl vs} a pseudo-colour of the continuum, similar to
U-V, has been plotted in figure \ref{fig3}.  In order to reproduce the
trend of HII galaxies, shifted to red colours at low values of
EW(H$_{\beta}$.)  due to the presence of a non ionizing population,
the contribution of the underlying population to the total continuum
must be higher than the contribution of the current burst which
dominates the emission line spectrum. This trend can not be reproduced
by SSPs or increasing metallicity or age separately \cite{mmm08a}, and
a strong attenuation is needed, as can be seen in the central panel,
to reproduce the whole range in EW(H$_{\beta}$) and colours
simultaneously.

\textbf{The time between burst ($\Delta$t) is a secondary parameter
which have an effect on the model similar to that of the attenuation.}
The reduction of the time between burst offsets the effect of
increasing the attenuation: colours of the models with shorter
inter-burst time are similar to those with soft attenuation (strong
bursts), and require an extra reddening to reproduce the effects of
the underlying non ionizing populations. However, the EW(H$_{\beta}$)
decreases from burst to burst while in the case of a soft attenuation
EW(H$_{\beta}$) maintains a high value . In order to reproduce the
correct behaviour of the HII galaxies, the inter-burst time can not be
less than 100 Myr for these models.

Besides this combination of values for the parameters, there are other possible combinations which could reproduce the observed features of HII galaxies \cite{mmm07, mmm08b, mmm08c}
%

\section{Conclusions}

We have made models which consist in instantaneous star formation
bursts spread along 13.2 Gyr. In order to reproduce the observable
characteristics of HII galaxies it is necessary to adjust three
principal parameters. With the \textbf{Initial efficiency} we can vary
the amount of gas involved in star bursts, which is going to lead the
SFR, the oxygen abundance, and the range of metallicity covered by the
emission lines produced by the ionized gas of the current burst of
SF. The \textbf{attenuation of the burst} sets the contribution of the
underlying continuum from the previous stellar generations born before
the stars of the current burst which dominates the spectrum. 
We can also change the \textbf{inter-burst time}, obtaining a similar effect in colours to the change in attenuation: decreasing this parameter, we can produce a larger contribution from the underlying continuum, as an increase in  the attenuation of the burst could do, but the effect in colours would be the same as decreasing the attenuation (making the burst stronger) thus making the spectrum bluer. At the same time, we have a larger contribution of the
underlying continuum, thus decreasing, even slightly, EW(H$_{\beta}$). Our method, based in these three parameter model, reproduce all observable characteristics of HII galaxies-abundances, colours and emission lines- at the same time.


\ack
This work has been partially supported by DGPTC grant AYA-2007-67965-C03-03 of the Spanish Ministry of Science and Innovation. Also, partial support from the Comunidad de Madrid under grant S-0505/ESP/000237 (ASTROCAM) is acknowledged.

\end{document}